\documentclass[a4paper,11pt]{article}

\usepackage{jcappub} 
\usepackage{mathtools}
\usepackage{hyperref}
\usepackage[T1]{fontenc} 
\usepackage{subcaption}
\usepackage{amsmath} 
\usepackage{amsfonts} 
\usepackage{amssymb} 
\usepackage{bm} 

\def\HI{H~{\sc i}\, }
\title{\boldmath Calibration requirement for Epoch of Reionization 21-cm signal  observation - III. Bias and variance in uGMRT ELAIS-N1  field power spectrum}

\def \iitbhu {Department of Physics, Indian Institute of Technology (Banaras Hindu University), Varanasi - 221005, India}
\def \dase {Department of Astronomy, Astrophysics and Space Engineering, Indian Institute of Technology, Indore, Madhya Pradesh, 453552, India}
\def \ncra {National Centre for Radio Astrophysics, Tata Institute of Fundamental Research, Pune 411007, India}
\def \iisc {Department of Physics, Indian Institute of Science, Bangalore 560012, India}
\def \mcgill {Department of Physics and McGill Space Institute, McGill University, Montreal, QC, Canada H3A 2T8}
\def \iitm {Cetre for Strings, Gravitation and Cosmology, Department of Physics, Indian Institute of Technology, Madras, Chennai 600036, India}
\def \knpg  {Department of Physics, K. N. Government P. G. College, Gyanpur, Bhadohi - 221304, India}

\author[a]{Saikat Gayen}
\author[b]{Rashmi Sagar}
\author[c,b]{Sarvesh Mangla}
\author[a]{Prasun Dutta}
\author[d]{Nirupam Roy}
\author[e]{Arnab Chakraborty}
\author[f]{Jais Kumar}
\author[b]{Abhirup Datta}
\author[g]{Samir Choudhuri}

\affiliation[a]{\iitbhu}
\affiliation[b]{\dase}
\affiliation[c]{\ncra}
\affiliation[d]{\iisc}
\affiliation[e]{\mcgill}
\affiliation[f]{\knpg}
\affiliation[g]{\iitm}
\emailAdd{saikatgayen.rs.phy22@itbhu.ac.in}
\emailAdd{phd2101121003@iiti.ac.in}
\emailAdd{phd1801121006@iiti.ac.in}
\emailAdd{pdutta.phy@itbhu.ac.in}
\emailAdd{nroy@iisc.ac.in }
\emailAdd{arnab.chakraborty2@mail.mcgill.ca}
\emailAdd{jaisk.rs.phy16@itbhu.ac.in}
\emailAdd{abhirup.datta@iiti.ac.in}
\emailAdd{samir@iitm.ac.in}

\abstract{Power spectrum of \HI 21-cm  radiation is one of  the promising probes to study large scale structure of the universe and understand galaxy formation and evolution. The presence of foregrounds, that are orders of magnitude larger  in the same   frequency range of the redshifted 21-cm signal  has been one of the largest observational challenges. The foreground contamination also hinders the calibration procedures and introduces residual calibration errors in the interferometric data. It has been shown that the calibration errors can introduce bias in the 21-cm power spectrum estimates and introduce additional systematics.  In this work, we assess the efficacy of 21-cm power spectrum estimation for  the uGMRT Band-3 observations of the ELAIS-N1 field. We first evaluate the statistics of the residual gain errors and perform additional  flagging based on these statistics.  We then use an analytical method to estimate the bias and variance in the power spectrum. We found that (a) the additional flagging based on calibration accuracy help reduce the bias and systematics in the power spectrum, (b)  the majority of the systematics at the lower angular scales, $\ell < 6000$,   are due to the residual gain errors, (c) for the uGMRT baseline configuration and system parameters, the standard deviation is always higher than the bias in the power spectrum estimates. Based on our analysis we observe that for an angular multipole of $\ell \sim3000$, $2000$  hours of `on source time' is required with the uGMRT to  detect redshifted 21-cm signal at $3-\sigma$  significance from a redshift of $2.55$. In this work we only consider the power spectrum measurement in the plane of the sky, an assessment  of residual gain statistics and its effect on multifrequency angular power spectrum estimation for the uGMRT and the SKA like telescopes will be presented in a companion paper.}


\begin{document}
\maketitle
\flushbottom

\section{Introduction}
\label{sec:intro}
Redshifted 21-cm signal from the large-scale distribution of neutral hydrogen (\HI) is one of the most promising probes to study the high-redshift Universe \cite{2001JApA...22...21B, 2010ARA&A..48..127M, 2013ExA....36..235M}. As this cosmological 21-cm signal is rather weak, a direct detection is difficult. Measurement of the two point correlation, like the correlation function or the angular power spectrum\cite{2007MNRAS.378..119D} of the redshifted 21-cm signal is considered as an effective probe of the large-scale structure\cite{2004ApJ...615....7M, 2004ApJ...608..622Z,2005MNRAS.356.1519B}. Observationally, radio interferometers are the key instruments to probe the redshifted 21-cm signal, where the visibility functions are directly measured. Correlating the visibilities at the nearby baselines gives estimates of the angular power spectrum of the signal \cite{2001JApA...22..293B}.

The 21-cm signal is present as a small component with relatively much higher  background signal in all low-frequency observations, and it is dominated by several orders of magnitude higher foreground radiation from other astrophysical sources \cite{2002ApJ...564..576D,2015ApJ...804...14T,2018IAUS..333..175P}.  Several techniques have been explored for foreground mitigation and  avoidance for detecting of 21-cm signal\cite{2009ApJ...695..183B, 2010MNRAS.409.1647J, 2011MNRAS.418.2584G, 2011MNRAS.413.1174P, 2013MNRAS.433..639P, 2011MNRAS.413.2103P, 2012MNRAS.423.2518C, 2013MNRAS.429..165C, 2012ApJ...749..164C, 2012MNRAS.419.3491L, 2012ApJ...752...80M, 2012ApJ...757..101T,2012ApJ...745..176V, 2013ApJ...769....5J, 2013ApJ...776....6T, 2014PhRvD..90b3018L, 2014PhRvD..90b3019L, 2014ApJ...788..106P, 2014ApJ...781...57S}.\\

Ghosh et al. (2012)\cite{2012MNRAS.426.3295G} estimate the power spectrum of sky intensity distribution using $10$ hours of the Giant Meter-wave Radio Telescope (GMRT)\footnote{See:  \url{http://www.gmrt.ncra.tifr.res.in}} data at 150 MHz frequency after subtraction of compact source component of foreground and find that at $k \sim 0.12$ $- $$1.2\  h\  $ Mpc$^{-1}$ the  upper limit of the power spectrum amplitude is $\sim 1000 mK^2$.  Paciga et al. (2013)\cite{2013MNRAS.433..639P} use 40 hours  of $150$ MHz observation of the  GMRT and report an upper limit of $248$ mK$^2$ at $ k \sim 0.5\  h Mpc^{-1}$. Barry et al. (2019)\cite{2019ApJ...884....1B},  with $21$ hours of the Marchison Widefield Array (MWA)\footnote{See: \url{https://www.mwatelescope.org}} observations,  report an upper limit of power spectral amplitude as $3900~mK^2$ at $k\sim 0.2\  h $ Mpc$^{-1}$ for a 21-cm observation from a redshift of $7.1$. A $141$ hours of observation with the Low Frequency Array (LOFAR)\footnote{See: \url{https://www.astron.nl/telescopes/lofar/}} gives an upper limit of $73$ mK $^2$ at the $k \sim 0.075\  h\ $ Mpc$^{-1}$ for 21-cm emission from the redshift of 9\cite{2020MNRAS.493.1662M}.

Presence of foreground in the 21-cm observations makes it a high dynamic range interferometric detection problem. Since the interferometer gains are  calibrated using  reference sky models, presence of foregrounds hinders the calibration accuracy. Gehlot et al. (2018)\cite{2018MNRAS.478.1484G}use  the LOFAR-LBA\footnote{Low Frequency Array - long baseline} to explore calibration errors such as gain errors, the effect of the polarized foregrounds, and ionospheric effects in power spectral analysis. Patil et al. (2016)\cite{2016MNRAS.463.4317P}   estimate the systematic bias due to the calibration error in context to the LOFAR-EoR\footnote{Low Frequency Array - Epoch of Reionization} experiments. Other literatures \cite{2015MNRAS.451.3709A,2016MNRAS.462.4482A,2018MNRAS.476.3051A,2015MNRAS.453..925V,2016MNRAS.458.3099V,2016RaSc...51..927M, 2022MNRAS.513..964M, 2023GeoRL..5003305M} have explored the effect of polarization leakage, ionospheric effects, etc. Barry et al. (2016)\cite{2016MNRAS.461.3135B} and Ewall-Wice et al. (2017)\cite{2017MNRAS.470.1849E} have discussed about effect of inaccurate models for sky-based self-calibration.  Redundancy calibration technique without a prior model of the sky is discussed in \cite{1982Natur.299..597N,1992ExA.....2..203W}. The redundancy calibration requires existence of redundant baselines, and hence is more effective for a certain type of array design. In this calibration, the gain solutions are independent of sky models, however, the overall amplitude and phase gradients have to be set with external information \cite{1992ExA.....2..203W,2010MNRAS.408.1029L}.  Non-redundancy in the baseline distribution results in spectral structure that contaminates EoR detections \cite{2010MNRAS.408.1029L}. Effect of antenna position offsets and beam variations on calibration solutions have been studied in \cite{2018AJ....156..285J, 2019MNRAS.487..537O}. Byrne et al. (2019)\cite{2019ApJ...875...70B} show that limitations of sky-based calibration result in a fundamental limit on the calibration accuracy and introduces additional spectral structure.\\

An important advancement in redshifted 21-cm power spectrum estimation is the introduction of the  visibility based Tapered Gridded Estimator (TGE)\cite{2016MNRAS.463.4093C}. In TGE the visibilities are gridded to gain computational efficiency. Additionally, the tapering of  the antenna response reduces the effect by bright sources from outside the field of view. Various versions of TGE have been implemented, like the TGE based multifrequency angular power spectrum estimator\cite{2019MNRAS.483.5694B}, the image based TGE\cite{2019MNRAS.483.3910C} etc. In absence of residual gain errors in interferometric calibration, the TGE based estimators avoid noise bias as well as effects due to incomplete baseline coverage.  

 Recently, Chakraborty et al (2019b)\cite{2019MNRAS.490..243C} presents uGMRT\footnote{The upgraded Giant Meter-wave Radio Telescope} Band-3 observations of  ELAIS-N1 field. They measure the power spectrum of  foreground emission for the redshifted 21-cm line in the frequency range of   $300 - 500$ MHz. Based on these observations,  \cite{2021ApJ...907L...7C} they note the upper limits on 21-cm power spectra as  $(58.87 \, \text{mK})^2$, $(61.49 \, \text{mK})^2$, $(60.89 \, \text{mK})^2$, and $(105.85 \, \text{mK})^2$ at the redshift  $z = 1.96$, $2.19$, $2.62$, and $3.58$, respectively for $k \sim 1 \, \text{Mpc}^{-1}$. These upper limits constrain the product of neutral \HI mass density ($\Omega_{H I}$) and \HI bias ($b_{H I}$) to the underlying dark matter density field as  $0.09$, $0.11$, $0.12$, and $0.24$ at the corresponding redshifts for $k \sim 1 \, \text{Mpc}^{-1}$. Pal et al (2022) \cite{2022MNRAS.516.2851P} use foreground avoidance technique to  estimate the power spectrum from uGMRT Band 3 observations of  ELAIS-N1 field at $z = 2.28$  and quote  $2-\sigma$ upper limit of $\Delta^2(k) \leq (133.97  \, \text{mK})^2$ for the 21-cm brightness temperature fluctuation at $k = 0.347 \, \text{Mpc}^{-1}$.  Elahi et al (2023a)\cite{2023MNRAS.520.2094E} investigate `Cross' Tapered Gridded Estimator (Cross-TGE), to estimate multi-frequency angular power spectrum (MAPS). In cross TGE, visibilities of two cross-polarisations (RR and LL) are correlated to reduce  systematics. Applying cross TGE in uGMRT Band 3 data at $z = 2.28$, they report a $2-\sigma$ upper limit  as  $\Delta^2(k) \sim (58.67)^2 \, \text{mK}^2$ at $k = 0.804 \, \text{Mpc}^{-1}$ constraining  $[\Omega_{\text{H I}} b_{\text{H I}}]^2 = 7.51 \times 10^{-4} \pm 1.47 \times 10^{-3}$. In a recent study Elahi et al. (2023b)\cite{2023MNRAS.525.3439E} use foreground removal to quote a  $2-\sigma$ upper limit on the 21-cm brightness temperature fluctuations as  $(18.07)^2 \, \text{mK}^2$ for $k = 0.247 \, \text{Mpc}^{-1}$ and  $[\Omega_{\text{H I}} b_{\text{H I}}]$ of $\leq 0.022$. 

Kumar et al. (2020)\cite{2020MNRAS.495.3683K} investigate  the effect of time-correlated residual gain errors in interferometric calibration by  simulating GMRT observation. They find that the residual gain errors introduce bias in the TGE based power spectrum estimates and  add to the systematics of the observation. Kumar et al. (2022)\cite{2022MNRAS.512..186K} explore the effect of time-correlated residual gain errors in presence of strong foregrounds and thermal noise analytically and provide the mathematical expressions for bias and variance in the power spectrum measurements with a few  assumptions.    \\
In this work we explore the validity of the assumptions and simplification made by Kumar et al. (2022)\cite{2022MNRAS.512..186K} using  uGMRT observations of the ELAIS-N1 field at $300-500$ MHz (data presented in Chakraborty et al. (2019b)\cite{2019MNRAS.490..243C}). We thoroughly investigate various characteristics of residual gain errors and assumptions made  in Kumar et al. (2022)\cite{2022MNRAS.512..186K}, to confirm the efficiency of their method and then use it to study  the  bias and excess variance in the power spectrum estimates of redshifted 21-cm signal. Rest of the paper is arranged in the following way. We briefly discuss the analytical estimates of the bias and variance in section~(2) and  salient features of the uGMR Band-3 observations of the ELAIS-N1 field in section~(3). In section~(4) we access the statistical characteristics of the residual gain errors. In section~(5) we estimate the bias and variance and then summarise the result and conclude.

\section{Analytical estimates of Bias and Variance of the TGE}
\label{baseline_pair_frac}
The interferometric gains and the residual gain errors vary with both time and frequency. Manifestation of the time correlated residual gain errors as a bias in the power spectrum estimate is shown in \cite{2020MNRAS.495.3683K}. It was observed by \cite{2022MNRAS.512..186K} that, with a few assumptions, the bias and excess variance in the time dependent residual gain errors can be analytically expressed for a known gain error model. In fact, this allows us to estimate the bias and variance in the power spectrum from a given observation, for which the gain error characteristics can be evaluated. In this work we use this analytical method to estimate the bias and variance in the 21-cm power spectrum from the ELAIS-N1 field observed with the uGMRT. Various results from this observations  are already  presented earlier {\cite{2019MNRAS.487.4102C,2019MNRAS.490..243C,2020MNRAS.494.3392C,2021ApJ...907L...7C,2022MNRAS.516.2851P,2023MNRAS.520.2094E,2023MNRAS.525.3439E} . In this section we briefly describe the steps to analytically estimate the bias and variance in the power spectrum for  known residual gain error characteristics, the detailed calculation can be found in \cite{2022MNRAS.512..186K}. We restrict ourself to only time dependence of the residual gains, the frequency dependence and its effects will be discussed in a companion paper. 

The recorded quantity, after calibration,  for each baseline $\vec{U_i}$ of an interferometer is termed as  the visibility $\tilde{V}(\vec{U}_i)$ and is related to the spatial coherence function of the sky signal $\tilde{V}^{S}(\vec{U}_i)$ as (see \cite{1996A&AS..117..137H, 1999ASPC..180.....T})
\begin{equation}
  \tilde{V}(\vec{U}_i) =  \langle \tilde{g}_A \tilde{g}^*_B \rangle \tilde{V}^S(\vec{U}_i) + \tilde{N}(\vec{U}_i),
  \label{eq:measurement} 
\end{equation}
where the measurement noise $\tilde{N}$  is expected to be Gaussian random, zero mean and not correlated across baselines. The standard deviation of the noise can be written as $\sigma_N = \frac{\text{SEFD}}{\sqrt{2\Delta \nu \Delta \tau}}$, where SEFD stands for the source equivalent flux density, $\Delta \nu$ and $\Delta \tau$ are the frequency and time range over which the visibilities are measured. As a part of standard procedure in an interferometric observation,  the  observed visibilities are calibrated for the antenna based gains. However, the estimation of the antenna based gains for the calibration process has limited accuracy, which results in non-unity residual gain in the calibrated visibilities. The quantities,  $\tilde{g}_{A}, \tilde{g}_{B}$ in eq~(\ref{eq:measurement})  are the residual antenna based gains
from the antennae A and B that defined the baseline $\vec{U}_i$. The residual gain from an individual antenna (e.g., antenna A) can be expressed as:
\begin{equation}
 \tilde{g}_A(t) = \left [ 1+\delta_{AR}(t) + i\delta_{AI}(t)\right ],
\end{equation}
where R and I  denotes  the real and imaginary parts. Note that for perfect calibration $\delta_{AR}(t) = \delta_{AI}(t) = 0$. Hence, in general,  the quantities $\delta$ has mean zero. We shall mention these quantities as the `residual gain error'  from hereon. The time correlation in residual gains  depends on the time delay $\tau$ and can be expressed by the  normalised two-point correlation function $ \xi_{A}(\tau)$ 
\begin{equation}
\xi_{A}(\tau) = \langle \delta_{A}(t) \delta_{A}(t+\tau) \rangle / \sigma_{A}^2,
 \label{eq:gainmod0}
\end{equation}
where $\sigma_{A}$ gives the standard deviations in residual gain errors for antenna $A$. 

Visibility correlation gives a direct measure of the power spectrum (\cite{2001JApA...22..293B}). This approach, along with variations introduced by \cite{2007MNRAS.378..119D, 2014MNRAS.445.4351C, 2016MNRAS.463.4093C, 2019MNRAS.483.3910C, 2019MNRAS.483.5694B}, among others, has been widely applied to estimate the angular power spectrum of the diffuse galactic foreground \cite{2012MNRAS.426.3295G, 2017NewA...57...94C, 2019MNRAS.487.4102C, 2020MNRAS.494.1936C} and the power spectrum of \HI distribution in nearby galaxies \cite{2009MNRAS.398..887D, 2013MNRAS.436L..49D, 2020MNRAS.496.1803N}. These studies assumes the interferometric calibration is perfect and  incorporate thermal-noise in  visibility estimate as well as  sample variance errors in power spectrum measurements, providing an assessments of power spectrum uncertainties.\\

Here we use the 2D power spectrum estimator discussed in \cite{2014MNRAS.445.4351C}, where visibilities are first gridded before power spectrum estimation. The size of the uv-grids $\Delta U$ is chosen to encompass an adequate number of baselines while ensuring all visibilities in the uv-grid remain correlated and is $\Delta U < \frac{1}{\pi \theta_0}$, where $\theta_0 = 0.6 \times \theta_{FWHM}$\footnote{FWHM: Full Width at Half Maxima}  of each antenna. Power spectrum estimation within each uv-grid  excludes visibility auto-correlations to significantly reduce noise bias. The contribution from each uv-grid within a specific annulus in $U = \mid \vec{U} \mid$ is combined, and the real part is used as the isotropic power spectrum estimate  for the baseline separation $U$. A rigorous expression for the uncertainties in the angular power spectrum estimated  with the TGE and the bare estimator can be found in \cite{2014MNRAS.445.4351C}. With sufficient accuracy, the  uncertainties in the angular power spectrum $C_{\ell} \mid _{HI}$ can be approximately expressed,  as \cite{2008MNRAS.385.2166A}
\begin{equation}
\sigma_{HI}^2 = \frac{C_{\ell} \mid _{HI}^2 }{N_G} + \frac{2 \sigma_N^2 C_{\ell} \mid _{HI} }{N_B N_d} + 2\frac{\sigma_N^4}{N_B N_d^2},
\label{eq:psvarng}
\end{equation}
where $N_G$ is the number of independent power spectrum estimates in a given annulus bin at $U$, $N_B$ is the total number of visibility pairs in the bin and $N_d$ is the number of days of observation. Note that we use the angular power spectrum $C_{\ell}$  expressed as a function of angular multipoles $\ell = 2 \pi U$  in the rest of the paper.\\

For a typical observation the characteristics of the gains, the signal etc has been well studied. Based on these, \cite{2022MNRAS.512..186K} considered the following simplifying assumptions.\begin{itemize}
    \item Residual gain errors are Gaussian random. 
    \item Residual  gain errors from different antennae are uncorrelated. 
    \item Real and imaginary parts of the residual gain errors  are uncorrelated.
\end{itemize}
Furthermore, it is expected that the residual  gain errors, redshifted 21-cm signal and foregrounds are statistically independent of each other. With these assumptions and additionally considering (a)  the gain statistics of all antenna are similar and (b) an adequate estimation of the foreground power spectrum ($C_{\ell} \mid _F$) is available, the bias and  variance in 21-cm power spectrum estimate can be given as \cite{2022MNRAS.512..186K}
\subsection*{Bias}
\begin{equation}
    \mathcal{B}_P=\left [(2n_1+n_3) \chi_2 +n_2\right]\frac{\Sigma_2}{N_d} C_{\ell} \mid _{F}
\label{eq:bias}
\end{equation}
\subsection*{variance}
\begin{eqnarray}
\sigma^2_{P}  = 
        \sigma^{2}_{HI} &+&
   \left [ \frac{4 \sigma_N^2 \Sigma_2  C_{\ell}\mid_F}{N_B N_d^2}\right] 
   +   8 \, \Sigma_4\,  \frac{C_{\ell}\mid_F^2}{N_G N_d^2} \\ \nonumber
   &+&  \left [(4n_1^2 + n_3^2) \, \chi_2^2+ n_2^2 \right ] \left [ 2 \Sigma_4 + \Sigma_2^2 \right ]  \frac{C_{\ell}\mid_F^2}{N_G N_d^2},
 \label{eq:variance}
\end{eqnarray}
with 
$$ \Sigma_2 = \sum_{\substack{AA = (RR, LL)\\ cc=(re, im)}}\sigma^2_{AA\_cc},  \ \ \Sigma_4 = \sum_{\substack{AA = (RR, LL)\\ cc=(re, im)}}\sigma^4_{AA\_cc},  \ \ 
 \chi_2 =  \sum_{\substack{AA = (RR, LL)\\ cc=(re, im)}} \chi_{AA\_cc} \sigma^2_{AA\_cc}  / \Sigma_2. $$    
Here we denote the standard deviation of real or imaginary parts of the residual gain for different polarisation ($RR$ and $LL$) as $\sigma_{RR\_re}$ and $\sigma_{RR\_im}$ or  $\sigma_{LL\_re}$ and $\sigma_{LL\_im}$ respectively. Note that, these values are considered to be similar for all the antennae.  The function
 $\chi(\ell)$ gives the integrated effect of the time-correlated residual gain errors over the uv-grids. If the correlation time of the residual gain errors $T_{corr}$ are larger than the integration time $\Delta \tau$ then
\begin{equation}
    \chi(\ell) = \frac{1}{T_D^2}\int_{\Delta \tau}^{T_D} (T_D-\tau)\xi(\tau)d\tau.
    \label{eq:chi}
\end{equation}
Here $T_D = \frac{\Delta U \times T_{24}}{\ell}$ is the time taken by baseline track of an antenna pair to cross a uv-grid of size $\Delta U$ at angular multipole $\ell$. $T_{24}$ corresponds to one sidereal day. Here the variance of the 21-cm power spectrum in presence of residual gain errors is indicated with $\sigma_P^2$, whereas, the quantity $\sigma_P$ gives the corresponding standard deviation. Latter can be interpreted as the effective uncertainty in the power spectrum estimates.

As we correlate the visibilities in a given baseline grid the pair of visibilities used can have different origins: 
\begin{itemize}
    \item \textbf{Type 1:} Correlation of visibilities measured by the same antenna pair at different times, expressed as \( \langle \tilde{V}_{AB}(t) \tilde{V}^*_{AB}(t') \rangle \).
    
    \item \textbf{Type 2:} Correlation of visibilities measured by antenna pairs with one common antenna, observed at the same time, given by \( \langle \tilde{V}_{AB}(t) \tilde{V}^*_{AC}(t) \rangle \).
    
    \item \textbf{Type 3:} Correlation of visibilities measured by antenna pairs with one common antenna, but at different times, denoted as \( \langle \tilde{V}_{AB}(t) \tilde{V}^*_{AC}(t') \rangle \).
    
    \item \textbf{Type 4:} Correlation of visibilities measured by antenna pairs with no common antenna, at any time, represented as \( \langle \tilde{V}_{AB}(t) \tilde{V}^*_{CD}(t') \rangle \).
\end{itemize}
The fraction of baseline pairs in a grid contributing from these different types are noted as $n_i$. Note that quantities $\mathcal{B}, \sigma_P, n_i, \chi, C_{\ell}$ depend on the angular multi-pole $\ell = 2 \pi U$.

The expression for bias and variance presented here uses direct error propagation from the observed visibilities with the simplifying assumptions stated earlier. We note that, the \textbf{Type 4}  baseline pairs does not contribute to either bias or excess variance  of the power spectrum. This is due to the fact, that for this type the correlation in visibilities comes from all four different antenna. This also suggests that if an interferometer is designed with dominant number of \textbf{Type 4}  baseline pairs, the residual gain errors will have minimum effect on the bias and excess variance. The bias is directly proportional to the Foreground power spectrum and the effective variance $\Sigma_2$ of the gain errors. The time correlation in gain errors depends on the baseline pair fractions of \textbf{Type 1, 3}. Since it is assumed that the antenna gain errors in different days of observations are not correlated, both bias and excess variance scales down with the number of days of observation. 

\section{uGMRT observations of ELAIS~N1 field}
\begin{table}[h!]
\centering
\begin{tabular}{cc}
\hline
Project code & 32\_120\\
Observation date & May 05, 06, 07 of 2017\\
  & June  27 of 2017\\
\hline
Bandwidth & 200 MHz\\
Frequency range & 300-500 MHz\\
Channels & 8192\\
Integration time & 2s\\
Correlations & RR RL LR LL\\
Total on-source time & 13 h (ELAIS N1)\\
Working antennas & 28\\
\hline
Flux Calibrator\\
Source & 3C286\\
Flux Density & 23 Jy\\
Source & 3C48\\
Flux Density & 42 Jy\\
\hline
Phase Calibrator\\
Source & J1549+506\\
Flux Density & 0.3 Jy\\
\hline
Target Field\\
Source & ELAIS N1 \\
\hline
\end{tabular}
\caption{ Observation details of the target field ELAIS N1 and calibrator sources for four observing sessions.}
\label{tab:observation}
\end{table}

We use the uGMRT observations towards ELAIS-N1  during GTAC cycle 32. The ELAIS-N1 field lies at high galactic latitudes ($l=86.95^{\circ}, b= +44.48^{\circ})$, and hence contribution of the Galactic synchrotron emission to foregrounds is relatively small for this patch of the sky. Needless to say that the foreground emission is still several orders of magnitude higher than the expected 21-cm signal. The observations were performed at  night time  to minimize ionospheric disturbance due to the Sun. The total observation time was  25 hours$k \sim 1 \, \text{Mpc}^{-1}$ spread over 4 days. For the present analysis, we  use the GMRT GWB data with 8192 channels over the frequency range 300-500 MHz, observed at 2~sec integration time. Details of the observations are given in Table~ \ref{tab:observation}. The raw visibilities are first initially calibrated following regular procedure after identification and rejection of the possible RFIs. A combined image was made from data of four days and then used for direction-independent self-calibration using a {\sc casa} \footnote{See: \url{https://casa.nrao.edu}; \cite{2007ASPC..376..127M}} based pipeline.  A final image of the field was made with a resolution of $4.6''\times 4.3''$ with a rms noise of $\tilde \sim 15 ~ \mu$Jy / beam. We use this final image along with the corresponding visibilities to estimate the residual gains through a self-calibration run in {\sc casa} with a {\sc solint} set to  $int$, which estimates the gains at an interval of $2.68$ sec. Note that, the gain tables are only obtained for the on-source time of observation. The gain tables, hence obtained, are used for further analysis. We also use the final visibilities, which are appropriately flagged, to estimate the baseline distributions and baseline pair fractions in our calculation. Further, all the sources with significant signal to noise in the final image are  identified and their effects are subtracted from the visibilities.  A detailed description of the data analysis, calibration and point source subtraction can be found in \cite{2019MNRAS.490..243C}, where they also estimate the DGSE\footnote{DGSE: Diffuse Galactic Synchrotron Emission} power spectrum at different sub-bands.

\section{Statistical Characteristics of the Residual Gain Errors}
In this section we evaluate the characteristics of the residual gain errors using the gain tables obtained earlier and access the validity of the assumptions discussed in section (2). We first show the results for the observation day of May 06 and then show a comparison of four days.

\subsection{Mean and standard deviation }
We extract the time series of the residual gains from the RR and LL polariazatins. We estimate the mean and standard deviation of these time series for each antenna. 
The standard deviation and mean of   the gain time series  are shown in the left panel of Figure \ref{fig:statis} for different antenna. We have used grey and black colors to denote RR and LL polarisations respectively. The real and imaginary parts of both polarizations are distinguished by circle and square markers. The y axis is scaled to show percentage variation with respect to unity. We found that the antenna 1, 13, 20, 21, 25, 27, 29 had gain standard deviation in excess of 8 percent. These antennas are not shown in either of the plots. We will discuss the effect of this additional flagging in a later section. We find that the mean of the standard deviations of  gain from different antenna are about $2.5-3.0 \%$ with about $1 \%$ variation across antennae. In this work, we use the mean values of the gain's standard deviation over all the antennae as representative values of $\sigma_{AA\_re}$ and $\sigma_{AA\_im}$ (where AA stands for RR and LL polarisation) for the particular polarization (LL/RR) and/or re/im parts. Note that, since we wish to use a simplified  analytical estimate of the bias and variance of the power spectrum here, we choose to use the mean of the gain standard deviations as mentioned above.  If the gain standard deviation vary drastically across the antenna a more accurate estimate of these quantities would require a thorough simulation.
 \begin{figure}[h!]
    \centering
    \includegraphics[width=0.47\textwidth]{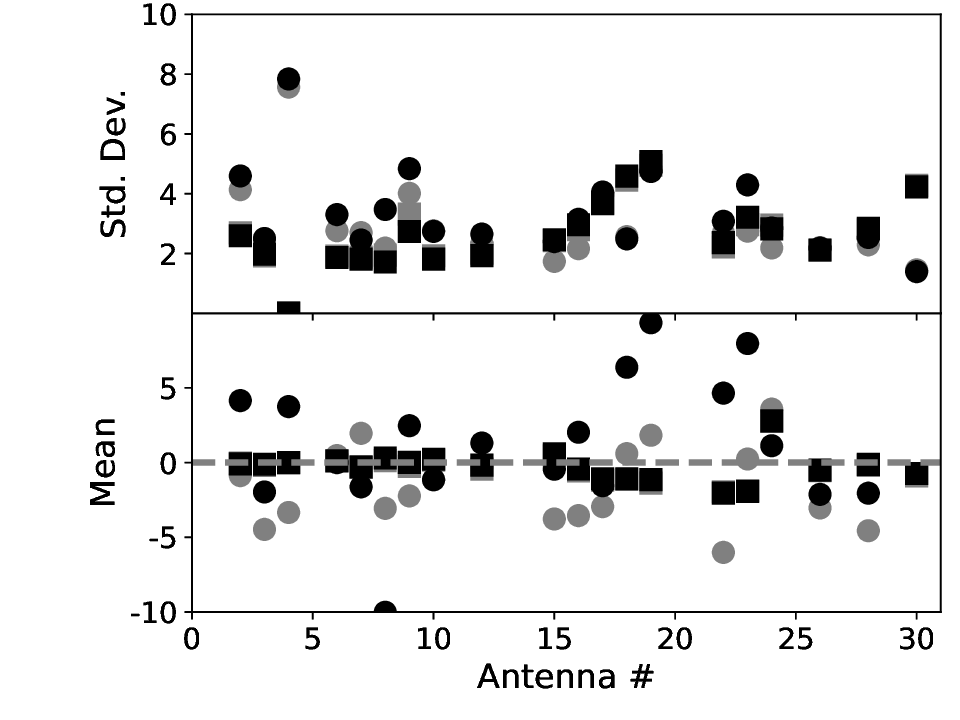}
    \includegraphics[width=0.47\textwidth]{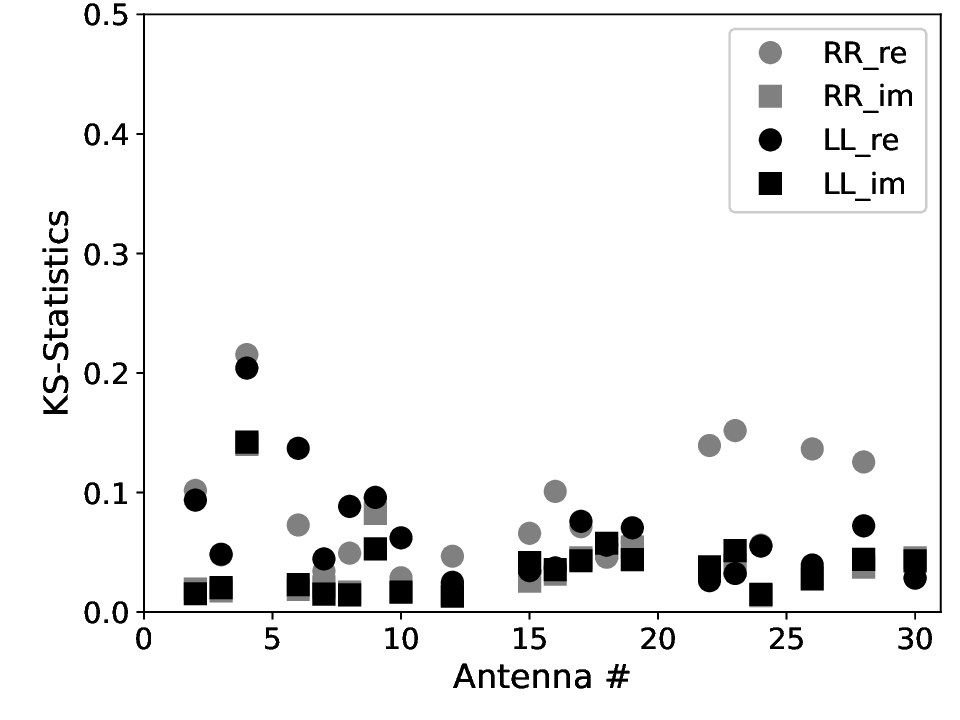}
     \caption{The standard deviation (left-upper), mean (left-lower) and KS statistics (right) of the residual gain errors for different antenna. We have used grey and black colors to denote RR and LL polarisations respectively. The real and imaginary parts of both polarizations are distinguished by circle and square markers. The y axes for the mean and standard deviation plots are scaled as percentage with respect to unity. We  show all the antenna that has a standard deviation of 8 percent and less.}
    \label{fig:statis}
\end{figure}

\subsection{Gaussian characteristics }
We use Kolmogorov-Smirnov test~\cite{1977ats..book.....K} on  the time series of the real and imaginary part of the gains from the RR and LL polariazatins for all the non-flagged antenna to access if they belong to Gaussian random distribution. The KS statistics for the gain time series are shown in the right panel of Figure \ref{fig:statis} for different antenna. Most of the antennae have a KS value less than 0.1 with only a few giving value to 0.2. This demonstrates, that there is only 10 percent chance that  most of the antennae not follow a Gaussian distribution. 

 \subsection{Time correlation}
 We  mentioned earlier that Kumar et al. (2022) \cite{2022MNRAS.512..186K} assumes that the residual gain errors are not correlated across different antenna, between real and imaginary parts or different polarisations. To test this, we estimate the normalized cross-structure function $S_2(\tau)\mid _{XY}$ of the residual gain errors defined as 
\begin{equation}
S_2(\tau)_{XY} = \langle  \left [ \delta_{X}(t) -  \delta_{Y}(t+\tau) \right ] ^2 \rangle / (\sigma_{X}\sigma_{Y}),
\end{equation}
where $X, Y$ corresponds to time series from different combinations of antenna, polarization etc.
The normalized structure function as defined above has a value of zero at zero lag ($\tau$), rises with increasing value of lag and remains $< 1$ since the time series is correlated. We define the correlation time to be the time at which  $S_2$ attains a value of $0.8$. To demonstrate this we plot the normalized self-structure function as a function of lag $\tau$  for the imaginary part of the LL polarization of residual gain error from antenna $16$ in Figure~\ref{fig:corrFunc} as points with error-bars. A best fit function is shown with black solid line. The function attains a value of $0.8$ at a lag of $\sim 49$ sec denoted as the correlation time $\tau_{corr}$ in the figure.
\begin{figure}[h!]
    \centering  
    \includegraphics[scale=0.47]{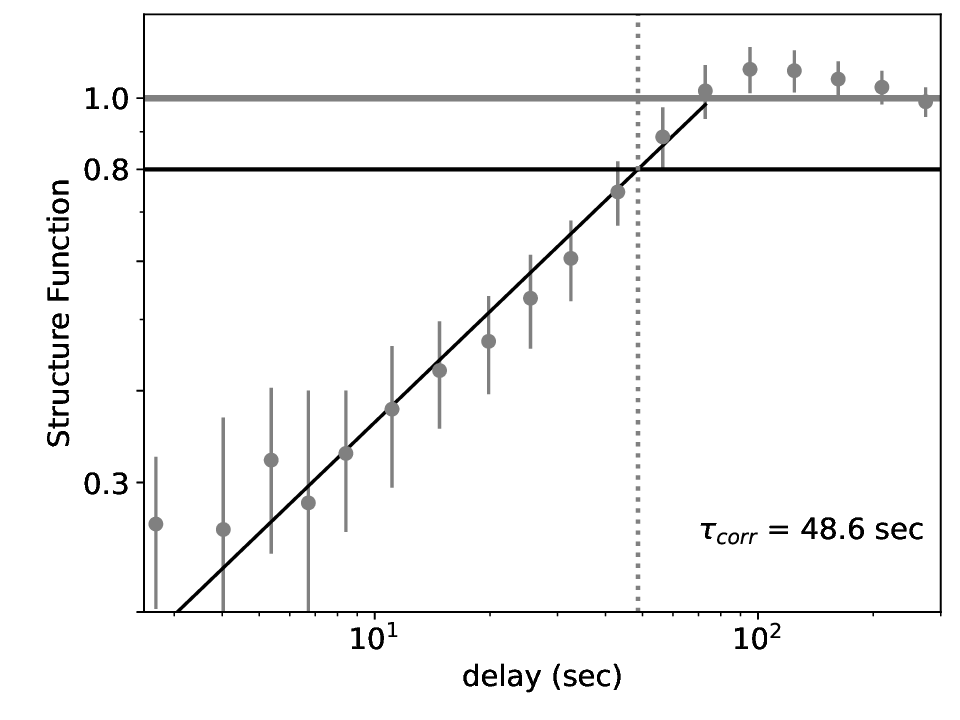}
    \includegraphics[scale=0.47]{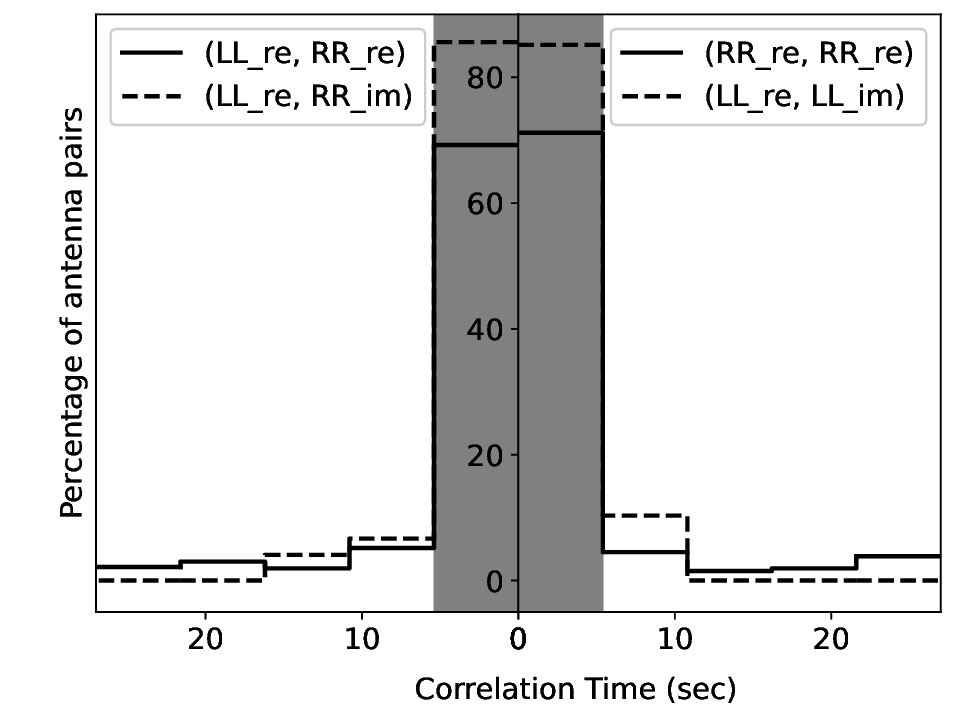}
    \caption{Left: An example of normalized structure function as a function of lag (grey points with error bars). A best fit power law to the estimated values between lag of $2$ and $70$ is shown as a black solid line. The horizontal solid black line represent a value of 0.8. The vertical dotted line denotes the correlation time of this particular time series. Right: Histogram of cross correlation time for different types of cross-correlations. The shaded region corresponds to twice the integration time for observation or the solution intervals for the gains.}
    \label{fig:corrFunc}
\end{figure}
To check any remaining time cross-correlation across different antennae/polarizations etc we find the correlation time for all possible cross-correlation. The integration time of the observation is $2.68$ sec and for insignificant cross-correlation we expect the correlation time to be within twice the integration time. We then divide these in bins of different combinations of polarization or real/imaginary parts. Black solid line in the right-hand panel of Figure~\ref{fig:corrFunc} show the histogram of correlation time from all possible different antenna pairs (LL\_re with RR\_re in left and RR\_re with RR\_re in right). Other combinations are also tested and few are shown here. We find that for all cases of cross correlation, the correlation time is less than twice the integration time for about $90 \%$ of the cases suggesting that the residual gain errors do not have significant cross-correlation.

Using the time series data of the residual gain errors of  each antenna, we estimate the time autocorrelation functions of both the polarizations. These numerical estimates of the autocorrelation function is then used to calculate the function $\chi(\ell)$ as given in eqn.~(\ref{eq:chi}). The angular field of view at $400$ MHz is about $2$ degrees, we choose a $\Delta U = 0.012$ k$\lambda$ ($\Delta U < 1/FoV$).  The estimated $\chi(\ell)$ from the imaginary part of the LL polarizations for all the antenna are shown with black dots in the left panel of Figure~\ref{fig:compfour}. The median values from all the antennae of the same function is shown with a black-dashed line. Clearly, there are not much significant variation across the antenna. The median values for all the antenna are plotted for both the RR and LL polarizations in the same plot. Note that, the $\chi(\ell)$ calculated from real parts of both the polarizations are smaller than their imaginary counterparts. At present we do not understand the reason for these systematic differences. Furthermore, the real and imaginary parts does not vary across the two polarizations. 
\begin{figure}[h!]
    \centering  
    \includegraphics[width=0.47\textwidth]{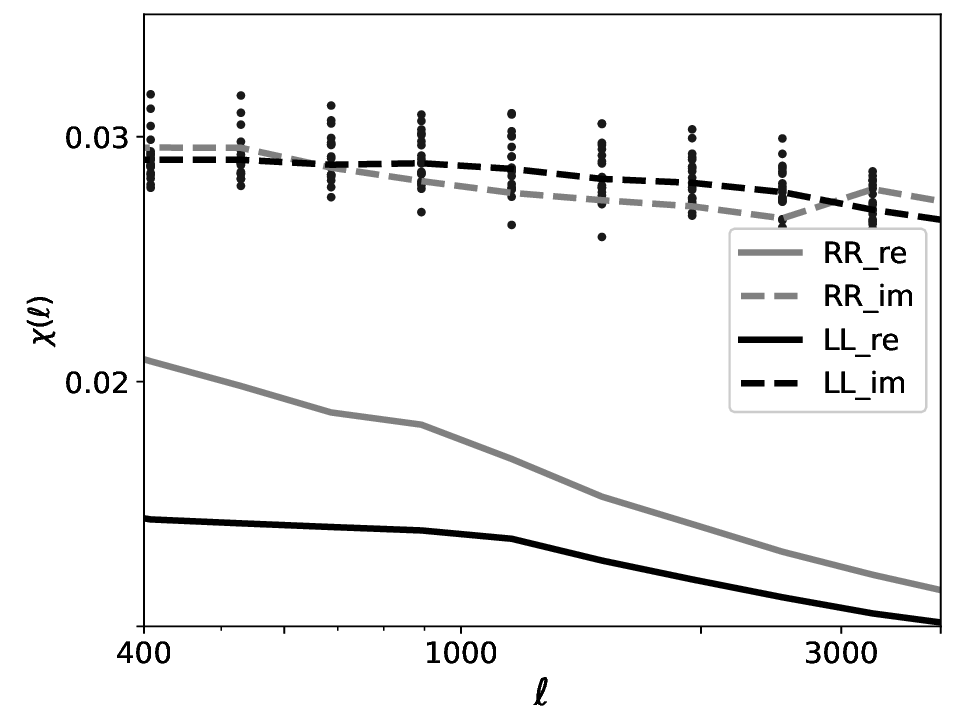}
    \includegraphics[scale=0.47]{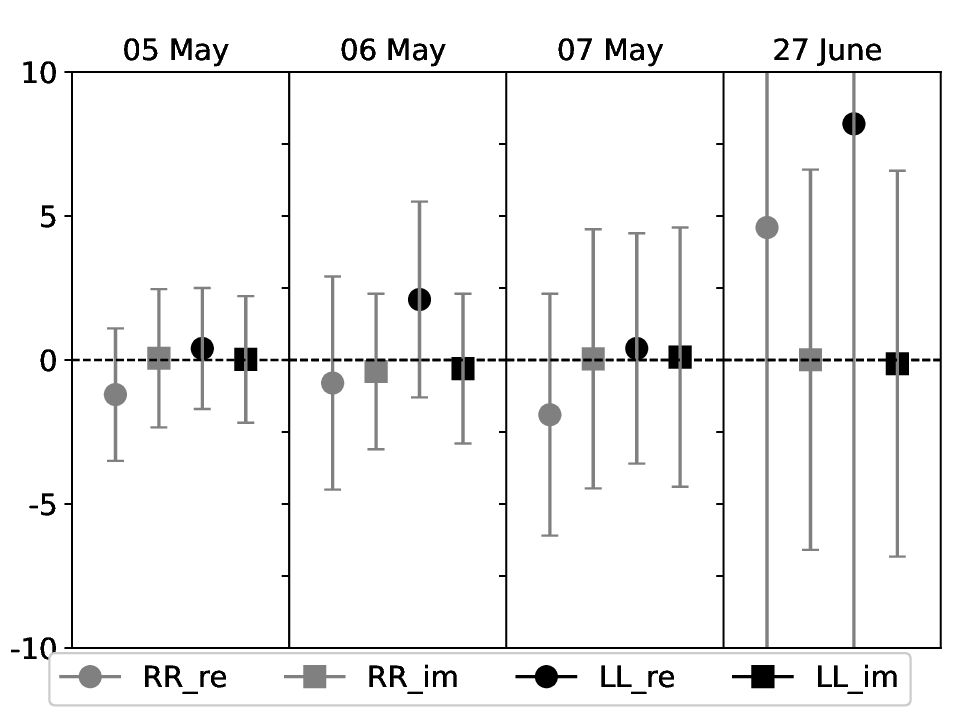}
    \caption{Left: We plot $\chi(\ell)$ with $\ell$, with grey color for RR and black for LL polarizations. The real parts of these polarizations are denoted with solid  and the imaginary part with dashed lines. We also show the $\chi(\ell)$ estimated for all the antenna for imaginary part of LL polarization with black dots. Right: Summary of gain characteristics of four observation days. Markers represents the mean of the residual gain errors from all the antennas. The mean of the standard deviation of all the antennas are shown as error bars.}
    \label{fig:compfour}
\end{figure}

\begin{table}[h!]
    \centering
    \begin{tabular}{ccccc}
        \hline
        Day & $\sigma_{RR\_re}$ & $\sigma_{RR\_im}$ & $\sigma_{LL\_re}$ & $\sigma_{LL\_im}$\\
        
        \hline
        May 05 & 2.3 & 2.4 & 2.1 & 2.2\\
        \hline
        May 06 & 3.0 & 2.7 & 3.1 & 2.6\\
        \hline
        May 07 & 4.2 & 4.5 & 4.0 & 4.5\\
        \hline
        June 27 & 16.1 & 6.6 & 18.4 & 6.7\\
        \hline   
    \end{tabular}
    \caption{Mean values of standard deviations of residual gain errors in percentages for all the used  antennae for four observing days.}
    \label{tab:SigmaSummary}
\end{table}
A comparison of statistics from four different days of observations is given in Table~\ref{tab:SigmaSummary} and is shown in the right panel of  Figure~\ref{fig:compfour}, where the markers represents the mean of the residual gain errors from all the antennas. The mean of the standard deviation of all the antennas are shown as error bars. We observe that the observation on June 27 has significantly high residual gain errors. The observations on May 05 and May 06 are better from the perspective of residual gain error.

\section{Results and Discussion}

\subsection{Baseline pair fractions}
We use  a grid size $0.012$ k$\lambda$  to estimate the baseline pair fractions. Left panel of Figure~\ref{fig:baseline} shows the baseline pair fractions $n_i$ from the observation of  May 06. As the baseline distribution mostly depends on the altitude/azimuth of the source, observations from the  other days show a similar pattern. We note that for these observation baseline pair fraction of Type~2 is completely absent.   The other three types of baseline pairs  have similar contributions in the $\ell$ range of our interest.
 \begin{figure}[h!]
    \centering
    \includegraphics[width=0.47\textwidth]{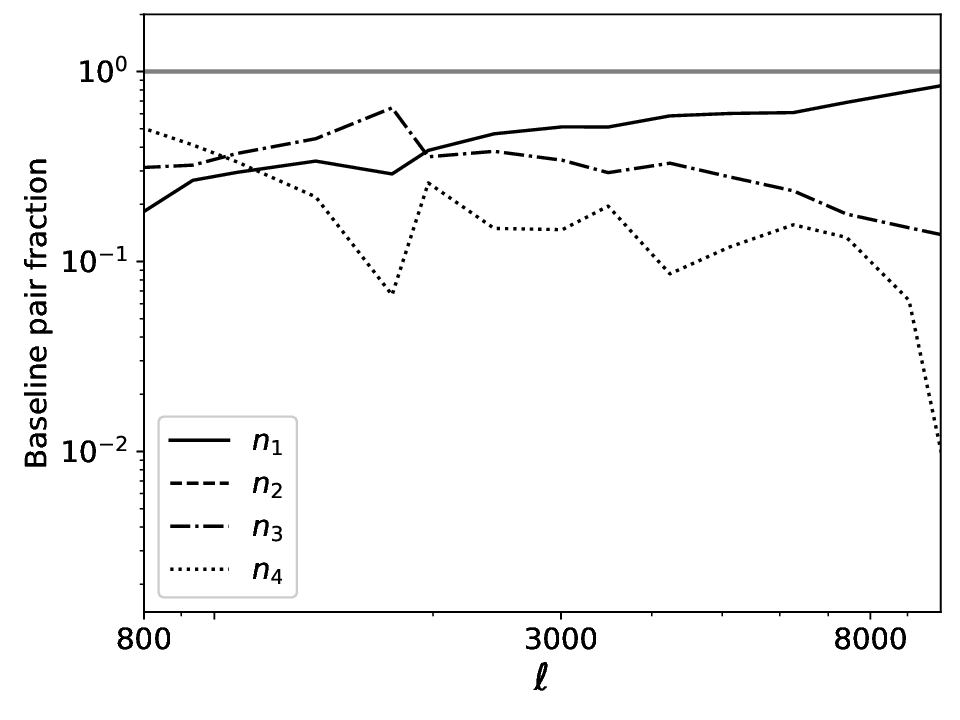}
    \includegraphics[width=0.47\textwidth]{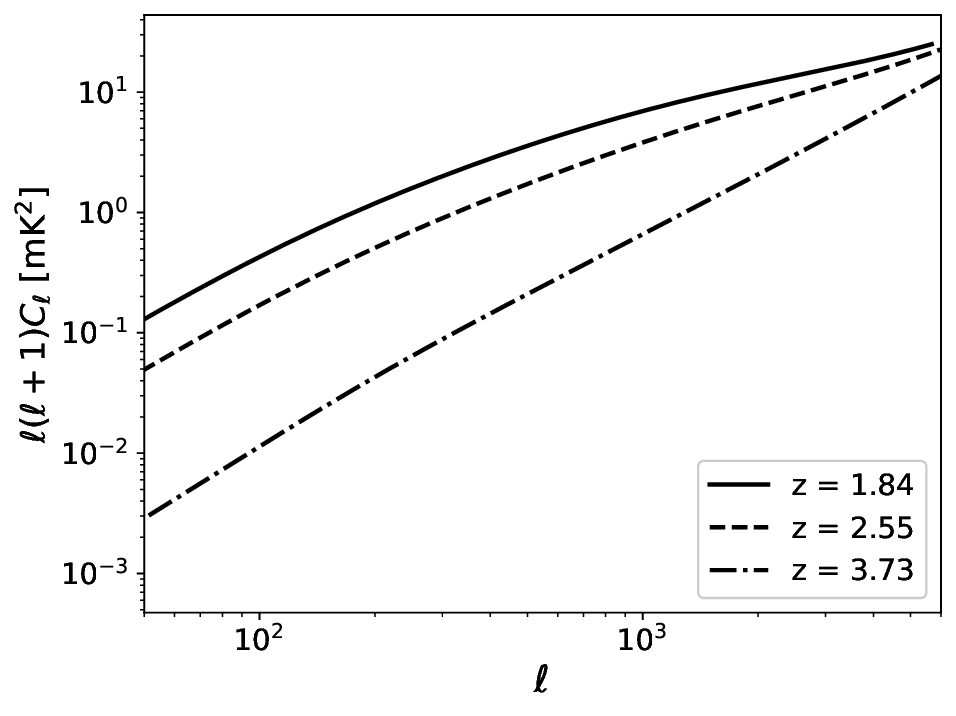}
    \caption{Left: Baseline pair fractions $n_i$ plotted against the angular multipole $\ell$ for the observation of May 06. Right: Expected  angular power spectrum of 21-cm signal is plotted with $\ell$ at redshift of $3.73, 2.55$ and $1.84$  corresponding to the observing frequencies  $300$, MHz $400$ MHz and $500$ MHz  with solid, dashed  and dash-dot black lines respectively. The 21-cm signal model is presented in eqn.~(5.1).}
    \label{fig:baseline}
\end{figure}

\subsection{Foreground model}
Chakraborty et al. (2019b) \cite{2019MNRAS.490..243C} present a thorough analysis of the foreground power spectrum along the ELAIS~N1 field with the same dataset used in this analysis. In an observation of redshifted 21-cm emission, the first step is to model and subtract the extragalactic compact sources.They used the calibrated data to make a deep image to model the point sources in the field and then subtract their contribution from the visibilities. The residual visibility then is dominated by the Diffuse Galactic Synchrotron Emission (DGSE). They then use the TGE to estimate the angular power spectrum of DGSE at eight frequency bins of $25$ MHz each. These power spectra are then fitted with  power laws of form $ C_{\ell}\mid _F  = A \left( \frac{1200}{\ell} \right)^\beta $. In our analysis, we use the estimated foreground  presented  in Figure 13 of \cite{2019MNRAS.490..243C} at a  frequency of $405$ MHz with  $25$ MHz  bandwidth. The best-fit  values of the parameters for this frequency are $A = 33.4 \pm 3.4$ and $\beta = 3.1 \pm 0.3$. The best fit power law was found for the angular multipole range of $600-3000$. The foreground does change across the observing bandwidth. However, for the scope of this paper, we use this as a representative foreground model across the $200$ MHz bandwidth of observation and evaluate the bias and variance of the power spectrum in the above angular multipole range. An observing frequency of $400$ MHz corresponds to \HI emission from a redshift of 2.55. The fiducial model for \HI signal and  the results presented here corresponds to this redshift.\\

\subsection{Model for power spectrum of 21-cm signal}
Post-reionization  21-cm angular power spectra has been modelled in \cite{2005MNRAS.356.1519B}, \cite{2014MNRAS.445.4351C} \cite{2016MNRAS.460.4310S} and \cite{2018MNRAS.476...96S}.  In these works, a particle-mess n-body simulation is used to evolve dark matter density to a model redshift. Different dark matter halo-finding algorithms are then used and the high density halos are populated with \HI. The \HI power spectrum is then estimated in 
redshift space including the redshift space distortion effects. Finally an analytical relation is presented to connect the \HI power spectrum with the dark matter power spectrum $P(k)$ through the scale dependent complex bias parameters $b(k), r $\footnote{Note that the bias $b(k)$ is a complex function. The quantity $r$ gives the ratio of the real part of the bias to the modulus of $b(k)$.}, redshift space distortion effects $\mu, \beta, D_{FoG}$ \cite{1987MNRAS.227....1K, 2018MNRAS.476...96S, 2019JCAP...09..024M} etc. We use the semi-analytical model of the 21-cm angular power spectrum as presented in Sarkar et al. (2016, 2018) \cite{2016MNRAS.460.4310S, 2018MNRAS.476...96S} to evaluate the \HI  angular power spectrum $C_{\ell}\mid _{HI}$ at the observation redshifts:
\begin{equation}
C_{\ell}\mid _{HI}  = \left(\frac{\partial B}{\partial T}\right)^{-2} \frac{1}{\pi r_c^2} \int_0^{\infty} {\rm d} k_{\parallel}  b(k)^2 \ \left [ 1 + 2 r \beta  \mu^2 + \beta^2 \mu^4 \right ]  D_{FoG} (k_{\parallel}, \sigma_v) P(k), 
\label{eq:HIPS}
\end{equation}
where $\mu$ is the cosine of  the angle between the wave vector $\vec{k} = [ \vec{k_{\perp}}, k_{\parallel}]$ and the line of sight direction. The quantity $r_c$ gives the comoving distance to the redshift of observation and the angular multipole $\ell = k_{\perp}r_c$. In this model the redshift space distortion at relatively smaller $k$ values arises from the factor $\mu$ and the redshift space distortion parameter $\beta$, whereas, at relatively smaller scales,  the finger of god effect is given by the function $D_{FoG}$, which varies with the component of the wave vector along the line of sight of observation, $k_{\parallel}$, and the pairwise velocity distortion $\sigma_v$ in a halo. The factor $\left(\frac{\partial B}{\partial T}\right)^{-2}$ converts the unit of the power spectrum from Jy$^{2}$ to Kelvin$^{2}$, where $B$ stands for the Plank function at temperature $T$.
We use this model to estimate the 21-cm angular power spectrum with  standard $\Lambda$CDM cosmology. The cosmological  parameters and matter power spectrum are taken from \cite{2016A&A...594A..13P} and  \cite{1994MNRAS.267.1020P}. We plot the variation of angular power spectrum with $\ell$ at redshift of $3.73, 2.55$ and $1.84$ in the right panel of Figure~\ref{fig:baseline}.  These redshifts corresponds to the centre and edges of the observational bandwidth of the data. 

 We use the expression given in eqn.~(2.5) and (2.6) to estimate the bias $\mathcal{B}_P$ and variance $\sigma_P^2$ of the angular power spectrum from observations of individual days. The representative numbers for $\sigma_{RR\_re}, \sigma_{RR\_im}, \sigma_{LL\_re}$ and $\sigma_{LL\_im}$  are taken from Table~\ref{tab:SigmaSummary}. The baseline pair fraction as well as the function $\chi$ are estimated for all four days. The results, presented henceforth, corresponds to a 21-cm signal at redshift of $2.55$.
  
 Left panel of Figue~\ref{fig:BiasVar} show the bias $\mathcal{B}_P$ and standard deviation $\sigma_P$  in the power spectrum as estimated for the observation day of May 06. The solid lines show the estimates of $\mathcal{B}_P$ as a function of the angular multipoles $\ell$, whereas the dashed lines show the $\sigma_P$. We show the fiducial estimate of the 21-cm power spectrum with grey dotted line. The dash-dot line show the expected standard deviation $\sigma_{HI}$ in the power spectrum in absence of residual gain errors. Note that the data was originally flagged for three antennae. The direct results from this is shown with grey lines (solid and dashed). Based on the poor gain statistics, we  flagged additional seven antennae. This additional flagging, in general,  reduces the number of baselines, however, also reduce the contribution of gain error from the antennas with poor gain statistics. The bias and standard deviation} in the power spectrum after the additional flagging  is shown with  black (solid and dashed) lines and are marked with "$\mid _F$". We find that for this observation, the additional flagging of antenna does not change the number of baseline pairs in the $\ell$ range plotted here and hence $\sigma_{HI}$ remains unchanged. We note that though the bias is not much sensitive to the additional flagging based on the gain statistics of the antenna, the standard deviation  improves. Clearly, this demonstrates an additional flagging based on the  statistics of the residual gains can improve the power spectrum estimation significantly. Henceforth, we shall be using the results with the  additionally flagging based on the gain statistics only. 
 
Dependence of bias and variance (and hence standard deviation) over the angular multipoles comes through the baseline pair fractions 
and the function $\chi$. We found that, the effect of the $\chi$ is rather small in the present observation data. The baseline pair fractions then introduces rather small bias in the power spectrum. In fact, the bias is less than the standard deviation in all angular multipole suggesting that the effect of gain error for this observation is to enhance the uncertainty in the power spectrum only. This excess variance, we believe, is the main systematics for the GMRT observations. We further observe, that the excess variance due to gain error reduces with angular multipoles and at relatively higher angular multipoles the residual gain error effects can be neglected. This nature of residual gain error is certainly observation and observatory specific and need to be assessed for each particular observation afresh.

Right panel of Figue~\ref{fig:BiasVar} show $\mathcal{B}_P \mid _F, \sigma_P \mid _F$  and $\sigma_{HI}$  for all four days of observations in different shades of grey. Clearly, the observation on May 05 has by far the best statistics, whereas the observation on June 27 is the worst. For the rest of the discussions, we shall consider the observation on May 05 as standard and use it as a representative case rather than combining the data from four day's observation. 

\begin{figure}[h!]
    \centering
    \includegraphics[scale=0.47]{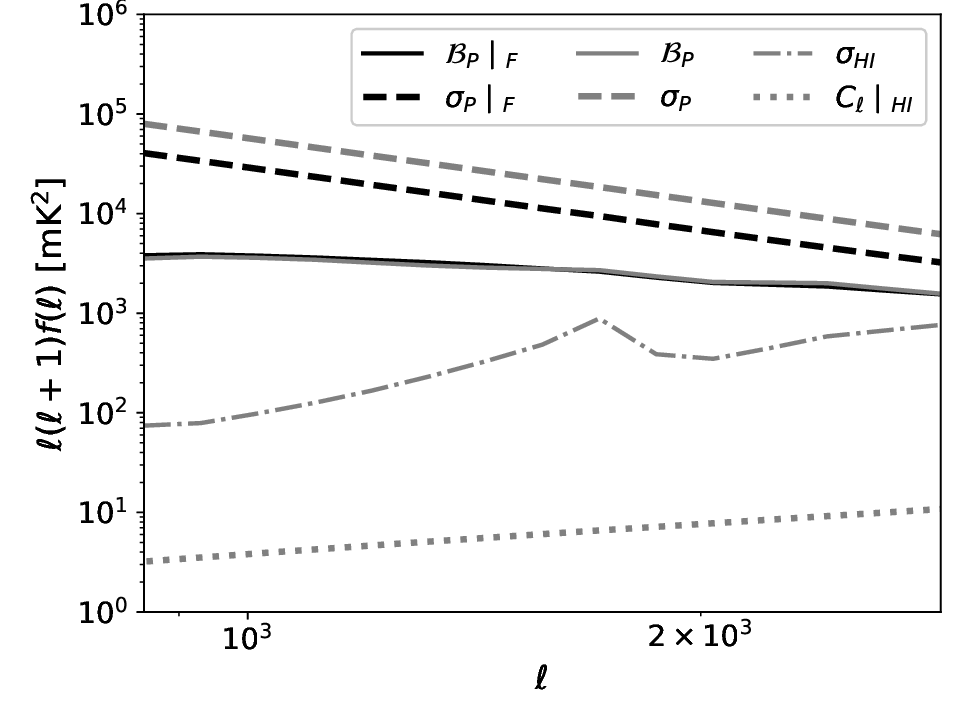}
    \includegraphics[scale=0.47]{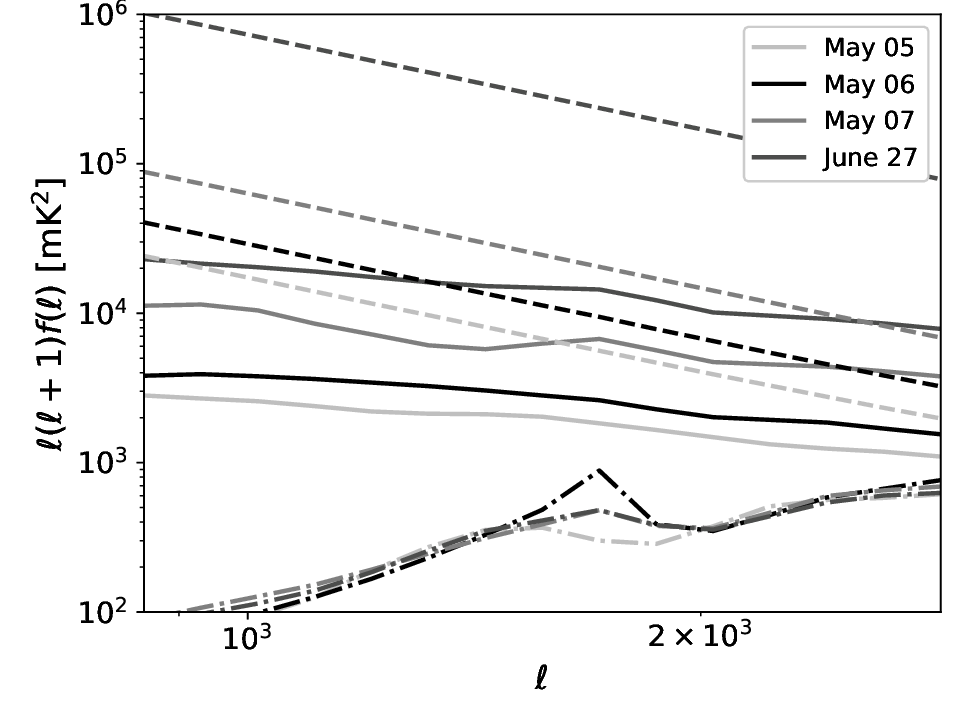}
     \caption{Left: Bias (solid lines) and  standard deviation (dashed lines) as a function of angular multipole for the observed data on May 06. Flagged bias and standard deviations  (marked with $\mid _F$ in legend) are plotted with black lines (solid and dashed respectively). The bias and standard deviation from the unflagged data are plotted with grey lines. The grey dash-dot line show the standard deviation in the power spectrum ($\sigma_{HI}$) when residual gain errors are zero. Note that in the multipole range plotted here the additional flagging has no visible effect on $\sigma_{HI}$.  Expected 21-cm signal power spectrum at redshift of $2.55$  is shown with the dotted grey line. All these quantities (as denoted collectively as $f(\ell)$ ) are scaled with  
$\ell(\ell+1)$ along the $y$-axis. Right: Bias (solid lines), standard deviations (dashed lines)  and $\sigma_{HI}$  (dash-dot lines) are shown for all four days of observation for comparison. }
    \label{fig:BiasVar}
\end{figure}

\begin{figure}[h!]
    \centering
    \includegraphics[scale=0.80]{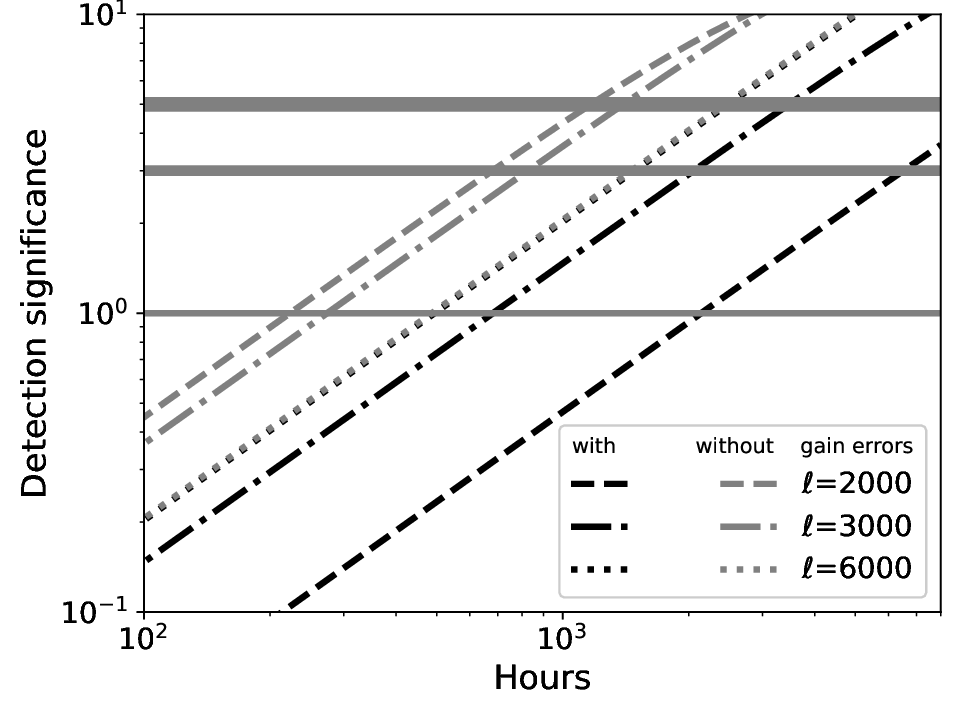}
    \caption{We plot for $\sigma$, $3\sigma$ and $5\sigma$ detection of 21-cm signal power spectrum for $\ell$ values of 2000, 3000 and 6000 with dashed, dashed-dot and dot-dot line respectively. Black  lines show the detection significance with  residual gain errors, whereas the grey lines show the detection significance in absence of (without) residual gain errors. $\sigma$, $3\sigma$ and $5\sigma$ line of variances are noted with solid grey lines.}
    \label{fig:Bias_3}
\end{figure}

Present attempts to observe the redshifted 21-cm signal is limited to uncertainties in the estimates of the power spectrum arising from different systematics. In this work, we accessed  the effect of time dependent residual gain errors in the uGMRT observation of ELAIS~N1 field and estimated the effective bias and standard deviation. For these observations the bias is found to be significantly lower than the standard deviation and hence can be ignored. We next estimate the detection significance of the redshifted 21-cm signal by defining it to be the ratio of the expected 21-cm signal $C_{\ell} \mid _{HI}$ to the corresponding standard deviations. 
Figure \ref{fig:Bias_3} show the detection significance of the 21-cm signal in Band-3, where we plot detection significance as a function of observing hours for fixed values of angular multipole $\ell$ by scaling the results from observation on May 05 analytically.  The   power spectrum is estimated with  the data from each full synthesis runs and then the  estimates from all the days are combined. We shall discuss how the result differ if the visibilities from all the days are used together to estimate the power spectrum shortly. The horizontal grey lines (of different thickness) mark detection significance of $1$, $3$ and $5$ respectively. The black curves (see legend) show the detection significance with  $\ell$ values of $2000, 3000$ and $6000$ in presence of residual gain errors. The corresponding grey curves gives the detection significance at same $\ell$ values (similar line types) when effect of gain errors are neglected. The effect of gain errors are more significant at lower values of $\ell$, we observe that at $\ell=6000$, for the uGMRT Band-3, the detection significance is dominated by the thermal noise (grey curves). The significance is also generally lower for a given time of observation when a lower $\ell$ are considered. We note that for $\ell=3000$, the uGMRT Band-3 will need about $2000$ hours of on source time to detect redshifted 21-cm signal at $3-\sigma$  significance from a redshift of $2.55$.  On the other hand for a $3-\sigma$ detection at $\ell=2000$ we need observation time of $6500$ hours and that for $\ell = 6000$ the time estimate is about $1500$ hours.
 
\subsection{Summary and discussion}
In this work, we evaluated the statistics of residual gain errors from uGMRT Band-3 observation of ELAIS-N1 during the GTAC Cycle~32. The observations were done over four days with about $3.5$ hours of on source time per day. We find that for the best day, most of the antenna the calibration is accurate to $5\%$ with a mean calibration accuracy of $2.3\%$, a rather good number in case of Band-3 of the uGMRT. We flagged an additional seven antennae where the calibration accuracy was relatively lower. The residual gain errors was found to follow a Gaussian statistics. The antenna gains showed  significant self correlation for many of the antenna, however the cross correlation of residual gains are mostly absent. We estimated the baseline pair fraction related to this observation, where we see that all but the baseline pair of Type~2 are not present for the uGMRT baseline configuration for the observation. These characteristics of the residual gain errors allowed us to use the analytical estimate of bias and variance in the power spectrum given in Kumar et al. (2022)\cite{2022MNRAS.512..186K}. This variance gives rise to the observed uncertainty in the \HI signal. Here we list our findings.


\begin{itemize}
\item We observe that the bias induced in the 21-cm power spectrum from the residual gain errors are orders of magnitude higher than the expected 21-cm signal, as well as the uncertainty in the power spectrum estimates in absence of gain error for our best observation day. However, the bias is also significantly lower compared to the uncertainty in the power spectrum when gain error effects are considered. Since the uncertainty and the bias in the power spectrum  scale similarly with the observation time, we expect the bias to be insignificant here. 

\item Excess uncertainty in the power spectrum estimate due to residual gain error manifests in the lower angular scales. We find that at angular scale of $\sim 6000$ or higher the uncertainty is dominated by the thermal noise only. 

\item We find excluding the antenna with poor gain characteristics significantly improve   variance in the power spectrum. This is an important observation, as this may help find analysis strategy to reduce the systematics coming from the residual gain errors.

\item Based on our analysis,   extrapolating the best observational case,  we expect with $2000$ hours of the uGMRT Band-3  observation, redshifted 21-cm signal from $z=2.55$ can be detected with $3-sigma$ significance at $\ell=3000$. 

\item We note that the excess variance is mostly from the contribution from the third and fourth terms from eqn.~(2.6), where the contributions scale as square of the variance in residual gain errors. Minimum variance in the residual gain error is a measure of the calibration accuracy, that can be achieved for a particular interferometer. This in turn depends on the ionospheric condition, the electronic gain stability  and the number of baselines used for a given interferometer. We found that for the uGMRT the standard deviation of the residual gain errors are  $\sim 2-2.5$ \%  (see Table 2). Clearly, with a factor of two improvement in the calibration estimates, that is for half the values of the residual gain standard deviation, the required time for 21-cm signal detection given above will reduce by a factor of four.
\end{itemize}

Note that, the time estimates for detection significance as given in the previous section assumes best observing conditions of the four days in our data. However, the gain characteristics of different observation days differ significantly, suggesting that in a real scenario, the actual observation time required to detect the 21-cm signal can be higher than quoted here. 
  Pal et al(2022)\cite{2022MNRAS.516.2851P} use a variant of TGE to estimate the 21-cm cylindrically averaged power spectrum and find a significant presence of systematics.  Elahi et al. (2023)\cite{2023MNRAS.520.2094E}  discuss the cross-TGE that can reduce much of the systematics  by cross correlating visibilities from RR with LL polarisation. In this work we observe that the antenna gains have significant time self correlations, however, the cross correlation in the residual gains from different antenna is almost absent. Manifestation of the time correlated antenna gains are in the function $\chi$. We find that, in our analysis,   effect of $\chi$ in the power spectrum uncertainty is only $3$ percent at maximum. If we use cross-TGE to estimate the power spectrum, the terms with $\chi$ would not contribute and for the present data, the bias in the power spectrum is absent. This suggests that cross-TGE can reduce the effect of residual gain errors. However, at lower multipoles, the effect of residual gain errors dominate over the thermal noise through the presence of $\sigma_{re/im}$ terms in the uncertainty estimate. 
 
 In this work, we adopt a strategy in estimation of 21-cm power spectrum by doing visibility correlation from each days observations separately and then combine the power spectrum from all  the observation days. This dramatically reduce the computational cost, however, one also is limited to lesser number of baseline pairs for visibility correlation. We find that if all the visibilities are used together to perform visibility correlation, then  the estimates of the bias and uncertainty of the power spectrum is modified. To understand this, we observe that the value of $\mathcal{B}_P, \sigma_P$ from one day of observation can be written with $N_d=1$ in eqn.~(\ref{eq:bias}, \ref{eq:variance}). If in a grid, the number of baselines is $N$, then the total number of baseline pairs in the bin would be $N_B =\  ^{N}C_2$. If we now correlate the visibilities of all observation days in a given grid, the number of baselines in a grid increases and hence there are much more baselines to perform cross-correlation of visibilities in a grid. This results in a change in the baseline pair fractions, which can be written as $n_i^{'} = \frac{N-1}{N N_d - 1} n_i$ for $i=1,2,3$ and be used in the expression for  $\mathcal{B}_P, \sigma_P$ with $N_d=1$. In the case where $N >> 1$, $n_i^{'} = \frac{1}{N_d} n_i$, the expression for $\mathcal{B}_P, \sigma_P$ mimics the eqn~(\ref{eq:bias}, \ref{eq:variance}). This suggest, that to a good approximation, our estimates of the bias and variance of the power spectrum can be used if all the visibilities from different days are correlated in a given grid for calculation of the power spectrum.
 
 In our analysis, we assumed that we have performed an adequate foreground removal and the foreground model is estimated from the data with sufficiently good accuracy. Inaccurate foreground subtraction as well as uncertainty in foreground estimates is expected to introduce additional uncertainty in the redshifted 21-cm power spectrum and has to be accessed separately. Using the errors in the power law foreground model, a  simple error propagation shows that for this frequency, the foreground model introduces an additional $2 \%$ increase in the power spectrum bias and an additional $5 \%$ uncertainty in the power spectrum variance.   
  
The upcoming radio interferometer, like the Square Kilometer Array Low (SKA-Low)\footnote{see \url{https://www.skao.int}},  is designed to have much better baseline coverage, whereas the baseline pair fractions of Type~4 is significantly higher. This is expected to drastically reduce the bias in the power spectrum estimate as well as lower the excess variance. Furthermore, better electronic design and baseline coverage would also reduce the excess variance in the power spectrum estimate. The analysis present in this paper discuss power spectrum estimate where visibility correlation is done in a single frequency channel only, where we use the representative frequency for the foreground as  $405$ MHz  with a thermal noise corresponding to a bandwidth of $200$ MHz. This approach certainly undermine the fact that the foreground changes significantly across the given bandwidth. A better approach would be to consider this variation of the foreground across the band along with the additional gain errors from the bandpass calibration and estimate the resultant excess bias and variance in the multifrequency angular power spectrum. We will present the effect of residual gain errors and observation time estimates for cylindrically and spherically averaged power spectrum for the SKA-low in a companion paper.

\section*{Acknowledgement}
We thank the staff of the GMRT who have made these observations possible. The GMRT is run by the National Centre for Radio Astrophysics of the Tata Institute of Fundamental Research. PD would like to acknowledge  research grant MATRICS (MTR/2023/000982) by SERB, India for funding a part of this project.

\bibliographystyle{JHEP}
\bibliography{references}

\end{document}